\documentclass[12pt]{article} 
\usepackage{latexsym} 
\begin{document} 
\input epsf 
\newcommand{\roughly}[1]%
        {\mathrel{\raise.4ex\hbox{$#1$\kern-.75em\lower1ex\hbox{$\sim$}}}} 
\newcommand\lsim{\roughly{<}} 
\newcommand\gsim{\roughly{>}} 
\newcommand\CE{{\cal E}} 
\newcommand\CL{{\cal L}} 
\newcommand\CM{{\cal M}} 
\newcommand\CO{{\cal O}} 
\newcommand\half{\frac{1}{2}} 
\newcommand\beq{\begin{eqnarray}} 
\newcommand\eeq{\end{eqnarray}} 
\newcommand\eqn[1]{\label{eq:#1}} 
\newcommand\intg{\int\,\sqrt{-g}\,} 
\newcommand\eq[1]{eq. (\ref{eq:#1})} 
\newcommand\meN[1]{\langle N \vert #1 \vert N \rangle} 
\newcommand\meNi[1]{\langle N_i \vert #1 \vert N_i \rangle} 
\newcommand\mep[1]{\langle p \vert #1 \vert p \rangle} 
\newcommand\men[1]{\langle n \vert #1 \vert n \rangle} 
\newcommand\mea[1]{\langle A \vert #1 \vert A \rangle} 
\def\Dsl{\,\raise.15ex \hbox{/}\mkern-12.8mu D} 
\newcommand\Tr{{\rm Tr\,}} 
\newcommand\eV{{\rm eV\,}} 
\newcommand\MeV{{\rm MeV\,}} 
\newcommand\TeV{{\rm TeV\,}} 
\thispagestyle{empty} 
\begin{titlepage} 
\begin{flushright} 
DOE/ER/41132-99-INT00 \\ 
 CALT-68-2291\\
\end{flushright} 
\vspace{1.0cm} 
\begin{center} 
{\LARGE \bf Couplings of a Light Dilaton and}\\  
\bigskip 
{\LARGE \bf Violations of the  Equivalence Principle}\\ 
~\\ 
\bigskip\bigskip 
{ David B. Kaplan$^a$ and Mark B. Wise$^b$} \\ 
~\\ 
\noindent 
{\it\ignorespaces 
           (a) Institute for Nuclear Theory, Box 1550\\ 
           University of Washington, Seattle, WA 98195-1550, USA   \\ 
           {\tt dbkaplan@phys.washington.edu}\\ 
\bigskip   (b) California Institute of Technology, Pasadena, CA 
91125 \\ 
{\tt   wise@theory.caltech.edu} 
}\bigskip 
\end{center} 
\vspace{2cm} 
\begin{abstract} 
Experimental discovery of the dilaton would provide strong evidence for string theory.
 A light dilaton could show up in current tests of the inverse 
square law for gravity at sub-millimeter distances. In the large extra  
dimension scenario, Kaluza-Klein excitations of the dilaton can also
contribute to the cooling of supernovae. In order to quantify these
effects we  
compute the couplings of a low energy dilaton to matter.
These predominantly arise from the 
fundamental dilaton coupling to the gluon field strength,  which 
receives a sizable enhancement from QCD scaling. We show that detection  
of the dilaton will give a direct measurement of the QCD coupling 
constant at the string scale. Particular attention 
is paid to the size of equivalence principle violating  dilaton couplings.  
\end{abstract} 
\vfill 
\today 
\end{titlepage} 
 
\section{Introduction} 
\label{sec:1}

String theory predicts the existence of a scalar particle called the 
dilaton, with prescribed tree-level couplings at the string scale. Its 
 discovery would be a major step toward experimentally 
validating string theory. It  
is believed that nonperturbative effects could generate a potential 
for the  
dilaton, but the mechanism and the form of this potential are 
presently  unknown. In particular, there is no theory for  the 
mass of the  
dilaton. Assuming dilaton mass generation is associated with 
supersymmetry breaking, a naive estimate would suggest that the dilaton 
mass is roughly $m\sim \Lambda_{SUSY}^2/M_{PL}$, where $\Lambda_{SUSY}$ 
is the supersymmetry breaking scale, and $M_{PL}$ is the Planck 
mass. $\Lambda_{SUSY}$ could be as low as a few 
TeV, corresponding to a dilaton Compton wavelength on the order of a 
millimeter.  Such a dilaton will mediate a measurable 
long range Yukawa force.  Currently 
there are several experiments testing for deviations from Newton's law  
of gravity at sub-millimeter distances 
\cite{Long:1998dk,Gundlach:1997re}.  In order to know what mass 
range for the dilaton they are sensitive to, it is necessary to know 
the dilaton's coupling to matter.   In the 
 large extra dimension scenarios, the dilaton can affect not only 
experimental tests of gravity, but also cooling rates of 
supernovae. 
In this note, we compute  
the leading contributions to the light dilaton  coupling to 
matter. Our analysis and results differ from earlier  
calculations of the dilaton coupling
\cite{Taylor:1988nw,Ellis:1989as}.

\section{Low Energy dilaton coupling to quarks and gluons} 
\label{sec:2}

We begin by computing the dilaton couplings assuming closed 
string theory above the string scale $\Lambda_s$,  and an effective 
field theory below $\Lambda_s$ consisting of the dilaton and  QCD with 
the three light flavors of quarks ($u,d,s$)  and an arbitrary number 
of heavy quarks, of which the charm quark is the lightest.  As 
we will discuss below, to leading order  
the only way that high energy physics enters the calculation (such as 
electroweak physics, supersymmetry, grand unification) is 
through the value of the strong coupling, 
$\alpha_3(\Lambda_s)$. Therefore the results and conclusions we derive in this 
section are actually independent of the particle content between the
weak scale and the string scale.  
 
Assuming that the dilaton $\phi$ is the only light modulus, and that
perturbation theory is valid at the string scale $\Lambda_s$,  the 
lagrangian for this system at $\Lambda_s$ is 
\beq  
\CL = \frac{1}{2}(\partial \phi)^2 - V(\phi) 
-\frac{1}{2}(1-\sqrt{2}\kappa 
\phi)\Tr G_{\mu\nu}G^{\mu\nu} + \sum_i\left[\bar q_i i \Dsl 
  q_i - \left(1+\frac{\kappa}{\sqrt{2}} \phi\right) m_i \bar q_i 
q_i\right] 
\eqn{lag} 
\eeq 
where $V(\phi)$ is the unknown potential for the dilaton (with minimum  
at $\phi=0$), $G_{\mu\nu}$ is 
the gluon field strength, and 
\beq 
\kappa=\sqrt{8\pi G_N} =\frac{\sqrt{8\pi}}{M_{PL}} = [2.43\times 
10^{18}\,{\rm GeV}]^{-1}\ . 
\eeq 
In \eq{lag} we have only kept those terms linear in $\kappa$.  In 
order to use this lagrangian to 
compute the coupling of a light dilaton to nucleons, we need to scale 
it down to the QCD scale $\sim 1$~GeV and match onto an effective theory 
of 
hadrons.  To do so, we employ the powerful techniques developed by 
Shifman, Vainshtein and Zakharov in 
ref. \cite{Shifman:1978zn}  to discuss the coupling of a light  
Higgs boson to nucleons (see also 
\cite{Voloshin:1987hp,Chivukula:1989ze,Chivukula:1989ds}).  We make 
use of the fact that both of the  
operators $m_i\bar q_i q_i$ and  $\partial_\mu S^\mu$ are 
renormalization 
group invariants, where  $\partial_\mu S^\mu$ is the divergence of the 
scale current: 
\beq 
\partial_\mu S^\mu =\frac{\beta_3}{g_3} \Tr 
G_{\mu\nu}G^{\mu\nu} + (1-\gamma_{m})\sum_i \, m_i \bar q_i q_i\ , 
\eqn{dscale} 
\eeq 
$\beta_3$ and $\gamma_m$ being the beta function for the QCD coupling 
$g_3$ 
and the mass anomalous dimension respectively. The coupling of the 
dilaton at the string scale  may be rewritten in terms of these 
operators as 
\beq 
\CL_{\phi}=\phi\frac{\kappa}{\sqrt{2}}\frac{g_3(\Lambda_s)}{\beta_3(\Lambda_s)}  
\left[ \partial_\mu S^\mu  
  - c \sum_i (m_i\bar q_i q_i)\right] 
\eqn{dlag} 
\eeq 
where 
\beq 
c = 
\left[1-\gamma_{m} +\frac{\beta_3}{g_3}\right]_{\mu=\Lambda_s}\  
. 
\eqn{cval} 
\eeq 
Note that $g_3(\Lambda_s)$ denotes the running QCD coupling evaluated 
at the string  scale. 
 
To compute the coupling of the dilaton to matter, we need to first 
integrate out the heavy quarks.  This is easily done to leading order 
in $\alpha_3$ by equating the divergence of the scale current in the 
theory renormalized  
just above the heaviest quark threshold, to the same quantity computed  
in the effective theory just below the charm quark mass.  One finds \cite{Shifman:1978zn} 
\beq 
\sum_{\rm heavy} m_i \bar q_i q_i = 
\left(1-\frac{b_0^{\, s}}{b_0^{\,\ell}}\right)\left(\partial_\mu S^\mu 
  -\sum_{\rm light} m_i \bar  q_i q_i\right) 
 +\CO\left(\alpha(m_c)\right)+\CO\left(\alpha(m_c)\Lambda_{QCD}^2/m_c^2\right) \ . 
\eqn{heavy} 
\eeq 
where the QCD $\beta$-function is given by $\beta=-b_0 
g^3/16\pi^2+\CO(g^5)$, where $b_0=(11-2/3 N_f)$ and $N_f$ is the 
number of quark flavors lighter than the renormalization scale. 
The quantity $b_0^{\, s}$ is computed at the string scale, including 
the heavy quark fields, 
while $b_0^{\,\ell}$ is computed in the effective theory with only the 
three  
light quarks.   Putting this 
altogether, the dilaton coupling may be written as 
\beq 
\CL_{\phi}=-\phi\frac{K \kappa}{\sqrt{2}}  \left[ 
  \partial_\mu S^\mu 
  -  \sum_{u,d,s} (m_i\bar q_i q_i)\right] +\ldots\ ,\qquad 
K=\frac{4\pi/b_0^{\,\ell}} {\alpha_3(\Lambda_s)} 
=\frac{4\pi/9}{\alpha_3(\Lambda_s)}  
\eqn{dlagf} 
\eeq 
where the ellipses refer to the $O(\alpha_3(m_c))$ and higher order 
corrections \footnote{This formula disagrees with two prior 
  calculations in the literature.   Our answer is similar to the 
  formula of Ellis et  
  al. \cite{Ellis:1989as}, but differs in two 
  ways:  First,  eq. (40) in that paper shows the 
  enhancement factor $K$ proportional to $1/b_0^{\, s}$ instead of 
  $1/b_0^{\,\ell}$ as in \eq{dlagf} above.  The difference arises from 
  our inclusion of the effects   
  of the heavy colored matter fields.  Second, we find that the terms responsible for  
  violations of the principle of equivalence receive the $K 
  =\CO(1/\alpha_3(\Lambda_s))$  
  enhancement, while they are found to be $\CO(1)$ in  
  \cite{Ellis:1989as}. The earlier work of Taylor and Veneziano 
  \cite{Taylor:1988nw} also found a logarithmic enhancement of the 
  dilaton coupling, but otherwise differs from our 
result.}.   
We have dropped the terms in \eq{cval} involving 
  $\gamma_m$ and  
$\beta_3/g_3$ evaluated at $\mu=\Lambda_s$, which  are subleading in 
$g_3(\Lambda_s)$, the strong coupling at the  
string scale. 
 
The result \eq{dlagf} is valid at leading order in  $\alpha_3$
provided the dilaton is the only light modulus (if there are several light moduli then
they can mix with each other). Our result \eq{dlagf} 
is independent
of assumptions about the nature of physics between the weak scale and
the string scale, other than that the gluon coupling remain
perturbative in this regime
\footnote{We explain in appendix~\ref{sec:7} why the inclusion of 
  electromagnetism does not change the form of \eq{dlagf} to the order
  we are working.}.  
Using the measured value of $\alpha_3(M_Z)$, the coupling $K$ can be computed in any particular model.  For 
example, in the standard model,  
$1/\alpha_3(M_{PL})\simeq 52$ and   
if $\Lambda_s=M_{PL}$ the enhancement factor is 
\beq 
K\simeq 73\qquad ({\rm Standard~~ Model,}\ \Lambda_s=M_{PL})\ . 
\eeq 
 In theories with additional heavy colored particles, or a lower
 string scale $\Lambda_s$,  the dilaton 
coupling will be weaker, because the QCD coupling will be larger at the
string scale.  For example in the minimal supersymmetric extension of
the standard model  
(MSSM) the scalar partners of the quarks and the
gluinos contribute to the beta  
function  above the weak scale, increasing the value of the strong coupling at the string scale. In 
this case $1/\alpha_3(M_{PL})=27$ and so  $\Lambda_s=M_{PL}$ implies an enhancement factor 
\beq 
K\simeq 38\qquad ({\rm MSSM,}\ \Lambda_s=M_{PL})\ . 
\eeq 
It is possible to have $K$ as small as
$\CO(1)$  if there is  a
large number of heavy colored states below the string scale, as occurs 
in many supersymmetric GUT models with $\Lambda_s=M_{PL}$. 
In any case, detecting the dilaton will give a direct 
measurement of the strong coupling constant at the string scale. 

We conclude this section with an aside about the relation between the 
dilaton mass and its coupling, imposed by naturalness. Naturalness is
simply the statement that the physical dilaton mass should be at least as large 
as the size of radiative corrections to the mass computed in
perturbation theory.  Radiative
corrections to the  dilaton mass  due to gluon loops are quadratically
divergent, and therefore sensitive to (unknown) high energy
physics. However the computable standard model contribution due to
gluon loops ought to serve as a lower bound on the radiative
corrections.  We estimate the standard model contribution with a cutoff of $\Lambda_{SM}=1~TeV$  to be
\beq
m_\phi\gsim  \left[\delta m_\phi \right]_{SM} &\simeq& \frac{1}{4\pi} \frac{\kappa}{\sqrt{2}}
\frac{\alpha_3(\Lambda_{SM})}{\alpha_3(\Lambda_s)}\Lambda_{SM}^2
\nonumber \\ \nonumber \\
&=&  2K\times 10^{-6}\,\eV\ .
\eeq
This lower bound is sufficiently weak that it does not constrain the
unexplored parameter space accessible to  experiments attempting to detect 
new long range forces.  However, if the scale of new physics is
higher, such as 
$\Lambda_{SM} \sim 10\,\TeV$, the naturalness constraint
becomes significant.

In any case, naturalness arguments are not
conclusive as one can imagine that the unnaturally small value 
of the cosmological constant  is somehow  related to
similarly unnatural properties of the dilaton. In fact, the currently favored
value of $\Omega_\Lambda\simeq 0.7$ implies an energy density $\sim
(10^{-3}\,\eV)^4$, making the exploration of (sub)-millimeter gravity 
especially intriguing \cite{Banks:1988je,Beane:1995sk,Beane:1997it}.

\section{The dilaton coupling to nucleons} 
\label{sec:3} 
 
To the order we are working, the  effective nucleon-dilaton coupling 
is given by  
\beq 
\CL_{N\phi} = \phi\left(y_p \bar p p + y_n \bar n n\right)\ ,  
\eeq 
where the   Yukawa couplings are given by the matrix element 
\beq 
y_i =-\frac{\kappa}{\sqrt{2}} K \langle N_i\vert 
\partial_\mu  
S^\mu  -\chi\vert N_i \rangle\ ,\qquad \chi \equiv (m_u\,\bar u u + m_d\, 
\bar d d + m_s\, \bar s s)\ , 
\eeq 
where $N_i$ is the nucleon doublet.  As the matrix element of the divergence of the scale current 
is 
 \beq 
 \langle N_i\vert  \partial_\mu 
S^\mu  \vert N_i \rangle = M_i\ , 
\eqn{mass} 
\eeq 
the Yukawa potential between two nucleons of types $i,j$ may be written 
as 
\beq 
\eqn{nucleonforce} 
V_{ij}(r) = -\frac{y_i y_j }{4\pi r}e^{-M_\phi r} = - K^2\, \frac{G_N \CM_i \CM_j}{ r}e^{-M_\phi r} 
\eeq 
where $K=(4\pi/9)/\alpha_3(\Lambda_s)$ is the enhancement factor in 
\eq{dlagf}, 
\beq  
\CM_p &=& M_p\left(1- \hat\chi_+ - \hat\chi_-\right)\ ,\nonumber\\ \nonumber\\ 
\CM_n &=& M_n\left(1-  \hat\chi_+ + \hat\chi_-\right)\ ,\eqn{nukes}
\\ \nonumber\\ 
 \hat\chi_\pm &=& \frac{1}{2}\left(\frac{\mep{\chi}}{M_p}\pm 
   \frac{\men{\chi}}{M_n}\right)\ . 
\nonumber
\eeq 
We compute $\chi_{\pm}$ to leading order in chiral perturbation theory
in appendix~\ref{sec:6}, finding
\beq 
\begin{array}{rcl}
\hat \chi_+ &=&
\ (2.7\pm 0.8)\times 10^{-1}\ ,\\  \\
\hat \chi_-  &=& -(1.5 \pm 0.5)\times 
10^{-3}\ , 
\end{array}
\eqn{chival}
\eeq 
where the errors quoted reflect an estimated $\sim 30\%$ violation of
$SU(3)$ symmetry.

In the
isospin symmetric 
limit $\hat\chi_-=0$  we see that the dilaton mediated force
between  nucleons obeys the principle of equivalence and is 
$
K^2 (1-\hat\chi_+)^2 
$
times stronger than gravity;  the factor  $K$ was
computed in \eq{dlagf}, while $(1-\hat\chi_+)^2\simeq 0.5$.   

The isospin violating term proportional to $\hat\chi_-$  contributes
to a force which violates the principle of
equivalence.  The difference between the  accelerations of a proton
and a neutron in a dilaton field is proportional to
\beq 
\Delta_{p,n}=\left(\frac{\CM_p}{M_p}-\frac{\CM_n}{M_n}\right) =-2  
\hat\chi_-\ , 
\eeq 
a $\sim 0.3\%$ effect relative to the common acceleration.
 
One might suppose that the inclusion of electromagnetism would give
 rise to comparable equivalence principle violating effects.  However,
the leading electromagnetic effects are already accounted for in
\eq{nukes} when 
the physical nucleon masses are used; additional electromagnetic
contributions are  suppressed by an additional power of $\alpha_{em}$
or $1/K$, as
discussed in appendix~\ref{sec:7}. 
 
\section{The dilaton coupling to atoms} 
\label{sec:4}

 
The Yukawa potential between two 
 atoms $i$ and $j$, from dilaton exchange, is 
 
\beq 
\eqn{atomforce} 
V_{ij}(r) = - K^2\, \frac{G_N \CM_i \CM_j}{ r}e^{-M_\phi r} 
\eeq 
where $K=(4\pi/9)/\alpha_3(\Lambda_s)$ is the enhancement factor in 
\eq{dlagf} and
\beq 
\begin{array}{rcl} 
\CM_i&=& \langle i |\partial_\mu S^\mu -\chi - m_e \bar e e |i\rangle
\\ \\
& =&
M_i-\langle i |\chi + m_e \bar e e |i\rangle \ ,  
\eqn{atme}
\end{array}
\eeq 
 where $M_i$ is the mass of atom $i$, and
we have included the electron 
coupling of the dilaton.  The mass of an atom with charge $Z$ and neutron number $N$ may be
written as
\beq
M = Z\, M_p + Z\, m_e +  N\, M_n -\CE\ ,
\eeq where $\CE$ is the binding energy,  $\CE\lsim A\times
(9\,\MeV)$, where $A=(Z+N)$ is the atomic number.  Thus we can write
the matrix element in \eq{atme}  as 
\beq 
\hskip-.2in
\langle i |\chi + m_e \bar e e |i\rangle   &=& \sum_{a=u,d,s,e} m_a \,
d M_i/d m_a \nonumber\\  \nonumber\\
&= & \left(Z_i M_p + N_i M_n\right)\hat\chi_+ + \left(Z_i
  M_p-N_i M_n\right)\hat\chi_- + Z_i m_e - \sum_a m_a \, 
\frac{d \CE}{d m_a} 
\eqn{chimat}
\\  \nonumber\\
&\simeq&
 M_i\left[\hat \chi_+ + \left({Z_i-N_i \over
      A_i}\right)\hat \chi_- +{m_e \over M_N}{Z_i \over A_i} - \sum_a m_a \, 
\frac{d (\CE/M_i)}{d m_a } \right]
 \nonumber
\eeq
In the last line we expanded to linear order in $\hat\chi_-$,
$m_e/M_i$, $(M_n-M_p)/M_N$ and $\CE/M_i$;  we have retained terms of order
$( \CE/M_i)\hat\chi_+$, which are of comparable magnitude to $Z_i m_e/M_i$
and $\hat \chi_-$.

We do not know how to compute the last term in \eq{chimat}, the
dependence of the binding energy on the quark and elecron masses.
It is expected to be dominated by the $m_s\bar s s$ operator, and  on dimensional grounds it 
should be of comparable magnitude to $( \CE/M_i)\hat\chi_+$.  We therefore define the parameters
$\xi_i$ by
\beq
\sum_a m_a \,
\frac{d (\CE/M_i)}{d m_a }\equiv \xi_i
\frac{\CE}{M_i}\hat\chi_+\ ,
\eqn{xidef}
\eeq
and expect  $|\xi_i|\sim
1$. 

Assembling these results we find that the parameter $\CM_i$ appearing
in the force law \eq{atomforce} is
\beq 
\eqn{atom} 
\CM_i \simeq M_i\left(1-\hat \chi_+ -\left({Z_i-N_i \over A_i}\right)\hat 
 \chi_--{m_e \over M_N}{Z_i \over A_i}-\xi_i
 \frac{\CE}{M_i}\hat\chi_+ \right)\ .
\eeq 
As for nucleons, we find that the equivalence principle conserving part
of the force is
$\sim K^2 (1-\hat \chi_+)^2 \sim 0.5 K^2$
times the gravitational force.  As discussed above, the numerical
value of  $K^2$ is unknown,
and could be $\CO(1) - \CO(10^3)$; it is  directly related to the
gluon coupling at the string scale through \eq{dlagf}.


\begin{figure}[t]
\beq
 \begin{array}{c||c|c|c|}
    & \left({Z_i-N_i \over A_i}\right)\hat \chi_- & {m_e \over
      M_N}{Z_i \over A_i} &\frac{\CE}{M_i} \hat\chi_+  \\[.1in] \hline
    \hline 
&&&\\[-.1in]
 {}^9_4{\rm  Be} & 1.7  & 2.42     & 19  \\[.1in] \hline
&&&\\[-.1in]
   {}^{27}_{13}{\rm  Al} & 0.56  &2.62    & 24  \\[.1in] \hline
&&&\\[-.1in]
    {}^{28}_{14}{\rm  Si} & 0  & 2.72     & 24  \\[.1in] \hline
&&&\\[-.1in]
    {}^{63}_{29}{\rm  Cu}& 1.2  & 2.51     & 25  \\[.1in] \hline
&&&\\[-.1in]
   {}^{208}_{\ 82}{\rm  Pb} & 3.2 & 2.15     & 23  \\[.1in] \hline
 \end{array}
\nonumber
\eqn{atoms}
\eeq
Table~1. {\it The last three terms in \eq{atom}, times  $10^4$, evaluated for various common
  isotopes;  we have  used the central values for $\hat \chi_{\pm}$ from first order
  chiral perturbation theory, \eq{chival}. The three terms are all
  small --- the first due to isospin violation, the second due to the
  small electron mass, and the third due to the smallness of nuclear
  binding energy.  Note that only the differences between entries for
  different isotopes, entering $\Delta_{ij}$ in \eq{deltadef}  are
  experimentally relevant for violation of the equivalence principle.} 
\end{figure}
 Violations of the equivalence principle in the
force between atoms $i$ and $j$ are  
proportional to 
\beq 
\Delta_{i,j}=\left({\CM_i \over M_i}-{\CM_j \over M_j}\right) 
\eqn{deltadef}
\eeq 
which gets contributions from the last three terms in \eq{atom}.  In
Table~1 we give the sizes for these equivalence principle violating
terms for several isotopes, given the chiral perturbation theory
calculation of $\hat\chi_+$ in \eq{chival}.
Note that by choosing to compare different pairs materials, it is in
principle possible to have the equivalence principle violating
parameter $\Delta_{ij}$ be selectively sensitive to each of the last three 
terms in \eq{atom}.

\section{Dilaton couplings with large extra dimensions} 
\label{sec:5} 
 
String theory has additional spatial dimensions beyond the three we have observed. In writing 
the Lagrangian in \eq{lag} we have assumed that these extra dimensions are 
small (size of order $1 /\Lambda_s$) and that only the particular
combination of higher dimensional metric components  
and higher dimensional dilaton field that appears as a factor scaling
the effective four dimensional action remains light.   
It is this combination that is the effective four 
dimensional dilaton 
field $\phi$ that occurs in \eq{lag}. Recently the possibility of having the string scale 
of order the weak scale has been explored
\cite{Antoniadis:1990ew,Lykken:1996fj,Arkani-Hamed:1998rs,Antoniadis:1998ig}.
The large effective four dimensional 
Planck mass arises because some of the additional extra dimensions have sizes much  
larger than $1/\Lambda_s$. In these scenerios   
enhancement of the low energy dilaton coupling to matter 
is different from that presented in \eq{dlagf} \cite{Dvali:2000hp}. This difference arises
for several reasons. Firstly  
to avoid experimental constraints the matter fields must be confined to a three-brane and 
at leading order in string perturbation theory the effective field theory arises from 
string amplitudes on the disk instead of the sphere. It is possible for there to 
be a dilaton mass term on the brane as well as in the bulk. Even with 
toroidal compactification and no warp factor a brane mass term 
gives rise to Kaluza-Klein excitations of the dilaton that are not just plane waves and 
a lowest mode that is not constant in the extra dimensions.  
This can dramatically influence the dilaton couplings to matter on the brane. In 
this section we assume that the dilaton couplings have 
the form one would deduce in the simplest possible case, where its lowest mass mode is 
constant in the extra dimensions and its Kaluza-Klein excitations are simply plane waves. 
  
With $n$ large extra dimensions the low energy 
dilaton couplings to quarks and gluons are given by \eq{dlagf} but the
enhancement factor is changed  
to
\beq 
\eqn{Kn} 
K_n={\sqrt{4+2n} \over 4}K 
=\frac{\pi \sqrt{4+2n}}{9 ~\alpha_3(\Lambda_s)}.  
\eeq 
Note that $K_n$ is a factor of $\sqrt{4+2n}/2$ larger than $K$ from
the normalization appropriate to rescaling the $4+n$ dimensional
metric to go from the string metric to the Einstein metric,  and an additional
factor of $2$ smaller   
because the couplings arise from string amplitudes calculated on the
disk. With two large extra dimensions and the  
string scale equal to $10~ \rm{TeV}$,  
 the enhancement factor in \eq{Kn} is $K_2 \simeq 15$. There is one more complication 
that effects the calculation of the force between matter from dilaton
exchange. The dilaton may not be a mass eigenstate,  
it might mix with components of the higher dimensional
metric. Assuming that this mixing angle $\theta$ is not  
small \footnote{For small $\theta$ terms that we neglected, which are not enhanced by $1/K_n$,  
 may be important.}   
, the force between matter 
from exchange of the lowest mass eigenstate is then given by
\eq{nucleonforce} (for nucleons) or \eq{atomforce} (for atoms)  
but with $K$ replaced by $K_n$ and a factor of $\cos^2\theta$ inserted. 
 
For two large extra dimensions an important constraint on the size of
the extra dimensions comes from the rate of supernova  
cooling associated with emission of Kaluza Klein excitations of the
graviton and  of a combination components 
of the higher dimensional metric which is also sometimes called the
dilaton. If the string theory dilaton is also light  
then it will contribute to this constraint. We can estimate the size of this effect using the  
recent results of Hanhart {\it et al.} \cite{Hanhart:2000er}. The
emissivity due to
Kaluza-Klein excitations of the  string theory dilaton is larger  
than  $\phi$ emission computed in ref.~\cite{Hanhart:2000er}  by a factor of
\beq 
\eqn{enhance} 
f={3(n+2)K_n^2 \over 4}. 
\eeq 
For $n=2$ and $\Lambda_s \simeq 10~{\rm TeV}$ this enhancement 
is $ f \simeq 550$. This results in a total emissivity approximately
8 times larger than computed in ref. \cite{Hanhart:2000er} for a
given size $R$ of the two flat extra dimensions.  The upper bound on
$R$  becomes stronger by $\sqrt{8}$, or
$R<2.5\times 10^{-4}\,{\rm mm}$.


\vskip2in
\centerline{\bf ACKNOWLEDGMENTS}
\bigskip

We thank J. Polchinski and M. Savage for useful comments. D.B.K. is supported in part by DOE grant
DE-FG03-00-ER-41132; M.B.W. is supported in part by  DOE grant
DE-FG03-92-ER-40701.
\vfill
\eject
\vskip2in
\appendix

\section{Computing $\CM_{p,n}$ in leading order chiral perturbation theory} 
 \label{sec:6}

The matrix elements $\hat \chi_\pm$ may be easily computed at leading order 
in chiral perturbation theory, using the 
 $SU(3)_L\times SU(3)_R$ baryon chiral lagrangian. Using the notation
 of ref. \cite{Nelson:1988sd},
the mass terms in this Lagrangian are,  
\beq 
\CL_m = -m_0\Tr\bar B B +2a_1 \Tr\bar B M B  + 2 a_2\Tr\bar B B M +2 
a_3 \Tr M \Tr \bar B B\ , 
\eeq 
where $M$ is the quark mass matrix ${\rm diag}\{m_u,m_d,m_s\}$, and 
$B$ is the baryon octet  
matrix.
The constants $m_0$ and $a_i$  times the quark masses can be 
determined in terms of the octet baryon masses, the quark mass 
ratios determined in the meson sector, and  the pion-nucleon 
``$\Sigma$-term'',  
\beq 
\Sigma_{\pi N}\equiv \left(\frac{m_u+m_d}{2}\right)\meN{\bar u u + \bar 
d d} 
\eeq 
extracted from $\pi-N$ scattering data.  The leading order result 
(i.e, to linear order in quark masses) is 
\beq 
\begin{array}{rcl}
\hat \chi_+ &=& \frac{1}{M_N}\meN{m_s\bar s s + \left(\frac{m_u + 
    m_d}{2}\right)(\bar u u + \bar d d)} \\ \\ 
 &=& 
\frac{1}{M_N}\left[\Sigma_{\pi N}\left(1+\frac{m_s}{m_u+m_d}\right) - 
  (M_\Xi +M_\Sigma-2M_N)  
  \left(\frac{m_s}{2m_s-m_u-m_d}\right)\right]\ , \\ \\ 
\hat \chi_- &=& \frac{1}{M_N}\mep{\left(\frac{m_u - 
      m_d}{2}\right)(\bar u u - \bar d d)}\\ \\ 
&=& 
\frac{1}{M_N}\left[ 
  -(M_\Xi-M_\Sigma)\left(\frac{m_d-m_u}{2m_s-m_u-m_d}\right)\right]  
\end{array}\eeq

The quark mass ratios have been determined from the measured pseudoscalar octet 
meson masses to be \cite{Leutwyler:1996qg} 
\beq 
 \frac{m_s}{m_u}= 34.4\pm 3.7\ ,\qquad \frac{m_s}{m_d}= 
18.9\pm 0.8 
\eeq 
while recent analyses of $\pi-N$ scattering  yields 
\cite{Gasser:1991ce,Buttiker:1999ap,Gasser:2000wv}  
\beq 
\Sigma_{\pi N} \simeq  45\,{\rm MeV}.\ 
\eeq 
These numbers, along with the measured hyperon masses yield 
\beq 
\begin{array}{rcl}
\hat \chi_+ &=& \frac{1}{2}\left( 
\frac{\CM_p}{M_p}+ \frac{\CM_n}{M_n} \right) = 
(2.7\pm 0.8)\times 10^{-1}\ ,\\  \\
\hat \chi_-  &=& \frac{1}{2}\left( 
\frac{\CM_p}{M_p}- \frac{\CM_n}{M_n} \right)= -(1.5 \pm 0.5)\times 
10^{-3}\ , 
\end{array}
\eqn{chivalapp}
\eeq 
where we have assigned (somewhat arbitrarily) a 30\% error estimate 
typical for $SU(3)$ violation. Note that in the standard model with
$\Lambda_s=M_{PL}$, $K\simeq 73$ and the 
dilaton force between nucleons is  
approximately $2.6\times 10^3$ times stronger the gravity, while the 
component violating the principle of equivalence  
is approximately $8$ times stronger than gravity  
\footnote{At leading order in chiral perturbation theory, $\chi_+$ is 
  dominated by the strange matrix element  
  $\mep {  m_s\, \bar s s}\simeq 218\,{\rm MeV}$.  This value is 
  somewhat controversial \cite{Jenkins:1992bs}. Reducing the  value of  
  $\mep { m_s\, \bar s s}$ has the 
  effect of boosting the overall strength of the force between 
  nucleons from dilaton exchange. For example, if $\mep{m_s\,\bar s 
    s}=0$, at short distance the dilaton force between nucleons is 
  enhance to $\sim 5\times 10^3$ stronger than gravity.}.

\section{Electromagnetic effects on dilaton coupling to nucleons} 
 \label{sec:7} 

The above analysis is apparently incomplete, as it neglects
electromagnetic effects.
One might think that the inclusion of  electromagnetism would change 
the equivalence violating potential by an amount comparable to that
due to the $(m_u-m_d)$ mass difference, as the two sources of isospin
violation contribute comparably to the  $n-p$ mass 
splitting.  While  true, such effects are already accounted for when the
physical proton and neutron masses (which include electromagnetic
contributions) are used in \eq{nukes}. 
All additional electromagnetic contributions 
to equivalence principle violation 
are relatively suppressed by either a factor of $1/K$ or
$\alpha_{em}$. 

Including electromagnetic interactions gives rise to a new coupling of
the dilaton at the string scale.  In addition 
to the terms in \eq{lag} there is now a term 
\beq  
\CL^{(em)}_\phi = -\frac{1}{4}(1-\sqrt{2}\kappa 
\phi)F_{\mu\nu}F^{\mu\nu}. 
\eqn{em}  
\eeq 
However when the dilaton couplings are scaled down from the string
scale its coupling to the electromagnetic field  
strength tensor, shown above, does not get enhanced by the large
factor $K$ and hence we can neglect the dilaton coupling  
in \eq{em}. The divergence of the scale current has a piece
proportional to the the square of the electromagnetic field  
strength tensor and so the low energy dilaton coupling is changed from
that in \eq{dlagf} to  
\beq 
\CL_{\phi}=\phi\frac{\kappa}{\sqrt{2}} K \left[ 
  \partial_\mu S^\mu 
  -  \chi-{\beta(e) \over 2e}F_{\mu \nu}F^{\mu \nu}\right] .  
\eeq 
The last term proportional to the square of the electromagnetic 
 field strength tensor 
has neutron and proton matrix elements of order $\alpha_{em}^2$ and
can be neglected.  
 The dominant electromagnetic 
effect is the electromagnetic correction to the matrix element of 
 the term in $\partial_\mu S^\mu$ proportional to the square of the gluon 
field strength tensor and it gives the contribution to the neutron and 
 proton masses of order $\alpha_{em} \Lambda_{QCD}$. This has already
 been included by using the  
full nucleon masses in \eq{nukes}.

\bibliography{dilaton} 
\bibliographystyle{h-physrev3.bst} 

\end{document}